\pdfoutput=1
 
\documentclass{sigchi-ext}

\usepackage[T1]{fontenc}
\usepackage{textcomp}
\usepackage[scaled=.92]{helvet} 
\usepackage{graphicx} 
\usepackage{balance}  
\usepackage{booktabs} 
\usepackage{ccicons}  
\usepackage{ragged2e} 

\usepackage{subfig}

\def\plaintitle{Towards Interaction Around Unmodified Camera-equipped Mobile Devices} 
\def\emptyauthor{}
\def\plainkeywords{around-device interaction; cross-device interaction; mobile interaction; wearable interaction; glasses; reflection; corneal imaging}

\title{Towards Interaction Around Unmodified Camera-equipped Mobile Devices}

\numberofauthors{4}
\author{%
    \alignauthor{%
        \textbf{Jens Grubert}\\
        \affaddr{Coburg University} \\
        \email{jg@jensgrubert.de} \\
         \vfill
    } 
    \alignauthor{%
        \textbf{Eyal Ofek}\\
        \textbf{Michel Pahud}\\    
        \affaddr{Microsoft Research}\\
        \email{eyalofek@microsoft.com} \\
        \email{mpahud@microsoft.com} \\
        \vfill
    }
     \alignauthor{%
     \vspace{0.5cm}
         \textbf{Matthias Kranz}\\
        \affaddr{University of Passau}\\
        \email{matthias.kranz@uni-passau.de} 
    } 
   \alignauthor{%
   \vspace{0.5cm}
     \textbf{Dieter Schmalstieg}\\
    \affaddr{Graz University of Technology}\\
    \email{schmalstieg@tugraz.at} 
    } 
}

\definecolor{linkColor}{RGB}{6,125,233}
\hypersetup{%
  pdftitle={\plaintitle},
  pdfauthor={\emptyauthor},
  pdfkeywords={\plainkeywords},
  bookmarksnumbered,
  pdfstartview={FitH},
  colorlinks,
  citecolor=black,
  filecolor=black,
  linkcolor=black,
  urlcolor=linkColor,
  breaklinks=true,
}

\begin{document}

\copyrightinfo{Copyright is held by the author/owner(s).
Presented at the Cross-Surface '16 workshop, in conjunction with ACM ISS '16.
November 6, Niagara Falls, Canada.}

\maketitle

\RaggedRight{} 

\begin{abstract}
  Around-device interaction promises to extend the input space of mobile and wearable devices beyond the common but restricted touchscreen. So far, most around-device interaction approaches rely on instrumenting the device or the environment with additional sensors. We believe, that the full potential of ordinary cameras, specifically user-facing cameras, which are integrated in most mobile devices today, are not used to their full potential, yet. We To this end, we present a novel approach for extending the input space around unmodified mobile devices using built-in front-facing cameras of unmodified handheld devices. Our approach estimates hand poses and gestures through reflections in sunglasses, ski goggles or visors. Thereby, GlassHands creates an enlarged input space, rivaling input reach on large touch displays. We discuss the idea, its limitations and future work.
\end{abstract}



\section{Introduction}
Handheld and wearable touch displays allow us to interact in a multitude of mobile contexts. However, shrinking device sizes, aiming at increased mobility~\cite{729537},  often sacrifice the interactive surface area. If devices shrink, while fingers stay the same, interaction may become inefficient
. Hence, there is a need for compensating for the lack of physical interaction area. 

One option is to decouple input and output area of interactive displays, using sensors to increase the input area around devices \cite{harrison2010appropriated}, extending it to near-by surfaces or to mid-air. Numerous research has sparked in the area of around-device interaction. So far, most research focused on equipping either mobiles~\cite{butler2008sidesight}, the environment~\cite{radle2014huddlelamp} or the user~\cite{chan2015cyclops,grubert2015multifi} with additional sensors. 
However, deployment of such  hardware modifications is hard. Market size considerations discourage application developers, which limits technology acceptance in the real-world~\cite{buxton2008long}.

We envision a future in which \textit{unmodified} mobile and wearable devices that are equipped with standard front-facing cameras allow for ample movements, including the environment around and to the sides of the device, without the need for equipping the device or the environment with additional sensing hardware. 

While there is a large body of work on around- and cross-device interaction most of them have the need for physically modifiying the device or equipping the environment with additional hardware \cite{grubertchallengens2016}. Song et. al \cite{song2014air} enabled in-air gestures using the front and back facing cameras of unmodified mobile devices.  However, their interaction space is limited to the field-of-view of the cameras, constraining the interaction space to two narrow cones in front and behind a device. Much of the interaction space around the mobile device, such as the areas to the sides of the device are not observed by these cameras. The closest work to ours is Surround See by Yang et al.~\cite{yang2013surround}. They modified the front-facing camera of a mobile phone with an omnidirectional lens, extending its field of view to $360^\circ$ horizontally. 
They showcased different application areas, including peripheral environment, object and activity detection, including hand gestures and pointing, but did not comment on the recognition accuracy. Their approach, just like ours, supports a large interactive space around the device. The need to add non-standard hardware to the phone limits the deployment of this technology. Furthermore, the $360^\circ$ lenses used by Yang et al. increase the size of the device thickness, making it hard to access and store a mobile device. In contrast, our approach only requires access to common and widely available apparels, which can result in a software only deployment to enable around-device interaction on millions of existing mobile devices.

In contrast to previous work, we propose to enrich the sensing capabilities of unmodified mobiles by everyday common apparels such as sunglasses or common reflective visors. 

Our interactive system, called \emph{GlassHands}, utilizes reflective glasses or visors to extend the field-of-view of front-facing cameras built into mobile devices, mimicking effects of catadioptric panorama cameras \cite{grubertGH2016}. Other reflective surfaces, such as ski goggles, diving masks or helmet visors, may enable gesture interaction with phones as well. This is of interest, when fine interaction with small screen is dangerous or impossible, for example, when the hands are covered with gloves. In fact, for some scenarios, such as skiing,  reflective visors are so common, that a software-only deployment may also be economically feasible. 

\def \teasersize {4cm}
\begin{figure*}
  \centering
  \subfloat{\includegraphics[height=\teasersize]{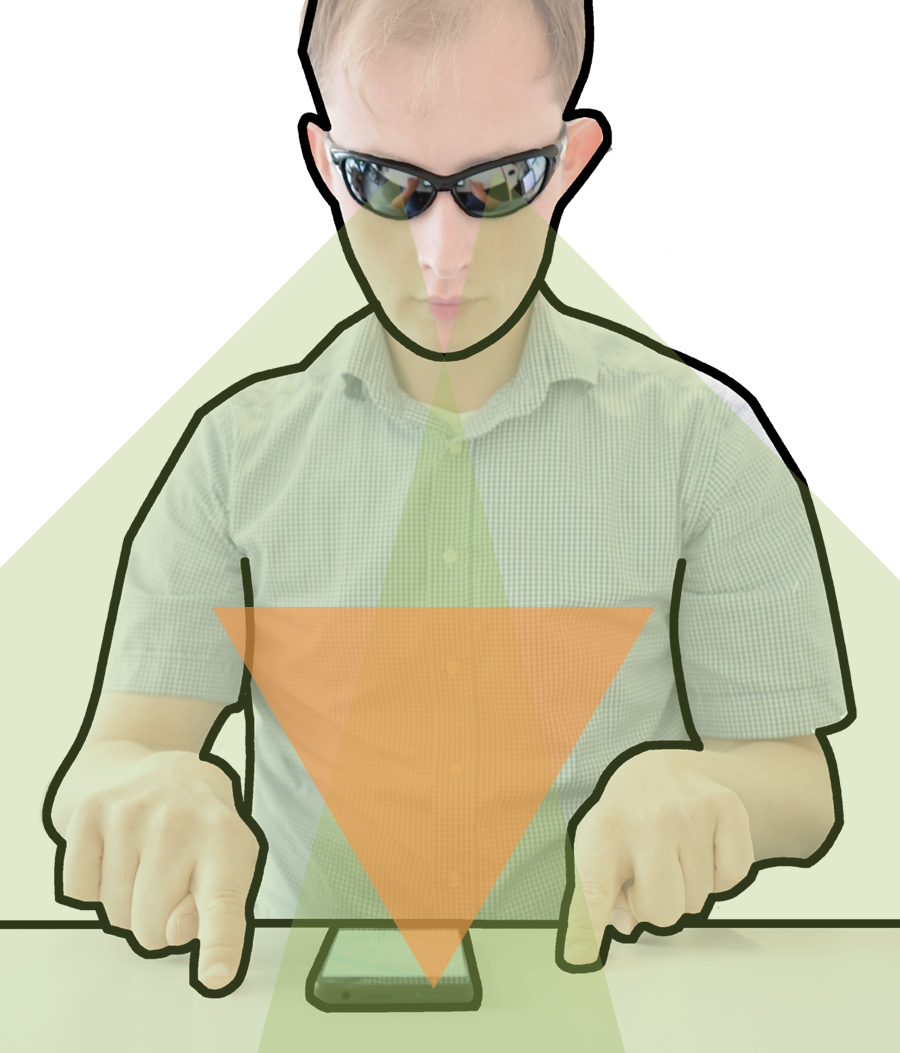}}~
\subfloat{\includegraphics[height=\teasersize]{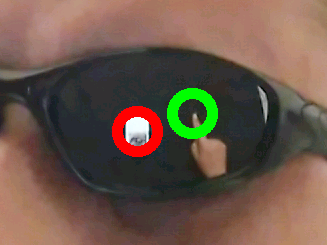}}~
\subfloat{\includegraphics[height=\teasersize]{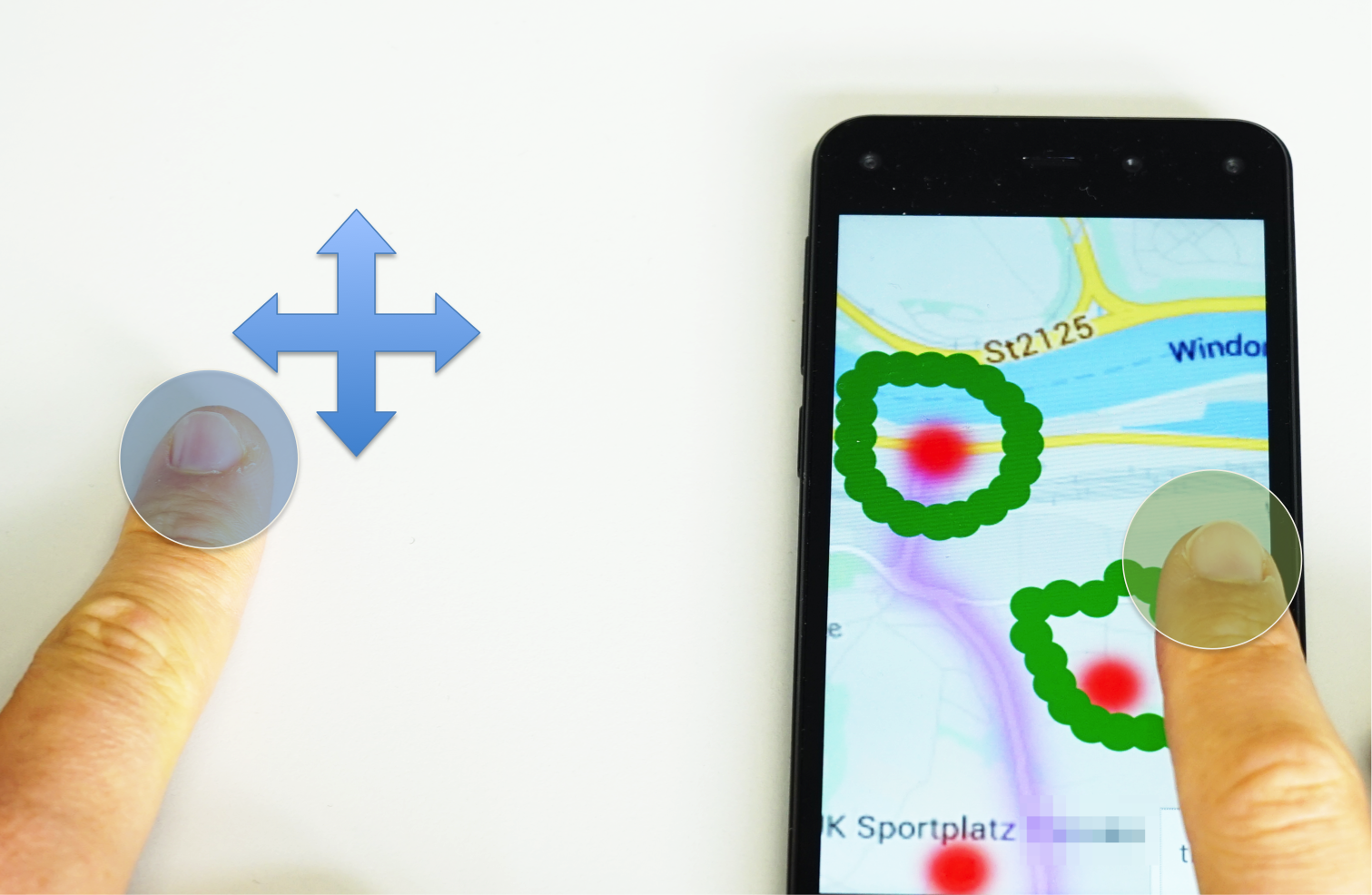}}
\caption{GlassHands extends the input space around mobile devices. Left: The narrow field-of-view of front facing cameras (orange) is extended through sunglasses reflections (green). Middle: Detected mobile phone (red) and finger tip (green) in glass area. Right: Users can continuously pan outside the display while simultaneously tracing over items on the device screen.}
\label{fig:teaser}
\end{figure*}

\newpage

\section{Around-Device Interaction using Reflective Glasses}

\begin{marginfigure}[-15pc]
  \begin{minipage}{\marginparwidth}
    \centering
    \includegraphics[width=0.9\marginparwidth]{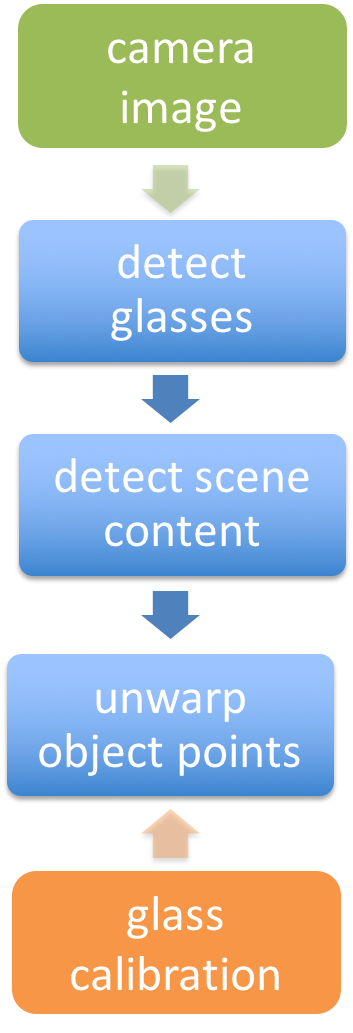}
    \caption{GlassHands processes image data (green) to determine the glasses region and scene content. Calibration data (orange) is used to transform phone and hand coordinates from the camera's image space to the display space of the phone.}~\label{fig:workflow}
  \end{minipage}
\end{marginfigure}


Off-phone interaction requires sensors that can observe the interaction, in our case, touch events on a surface and in-air events around the phone. However, the unmodified phone does not have any such sensing ability. GlassHands mimics a virtual external point-of-view (POV) that captures large areas of the workspace, by using the existing camera to observe a reflection from surface that lies in front of the camera. The reflection will contain the phone itself, the surface around the phone, and the user's hands.


Such a reflectors may be part of the environment, mounted on a ceiling or above a workstation. In this work, we look at \emph{wearable} reflectors such as glasses or visors. Being worn in front of the user eyes, these reflectors are naturally positioned to face the phone. In many useful scenarios where normal touch interaction is difficult, for example, when the user has to wear gloves, or dangerous, such as skiing, bike riding, diving, manufacturing and more, such eye-wear is already common and can be leveraged.

To use the reflection image as an input modality,  we need to be able to detect the reflected image in the camera image and extract the relative location of objects that need to be sensed. Specifically, we must detect the position of the user's hands next to the phone and, finally, map this information to world coordinates.

The components of the system encompass detection of the user head and the area of the user's glasses where the reflection is visible, detection of the phone and the user hands in the reflection, and translation of all these locations to metric world coordinates around the phone. Finally, the implementation of a touch interface requires the ability to detect when the user's hands touch the surface.

We have implemented the system on an Amazon fire phone using OpenCV for image processing and HTML5 and JavaScript for application development. We note, that while our system was implemented on an Amazon fire phone, it can be employed on other commodity smartphones as well. The system workflow is summarized in Figure \ref{fig:workflow}.

\section{Applications}
We demonstrate the potential of our concept by implementing various application prototypes, which are described next.

\begin{marginfigure}[5pc]
  \begin{minipage}{\marginparwidth}
    \centering
    \includegraphics[width=0.9\marginparwidth]{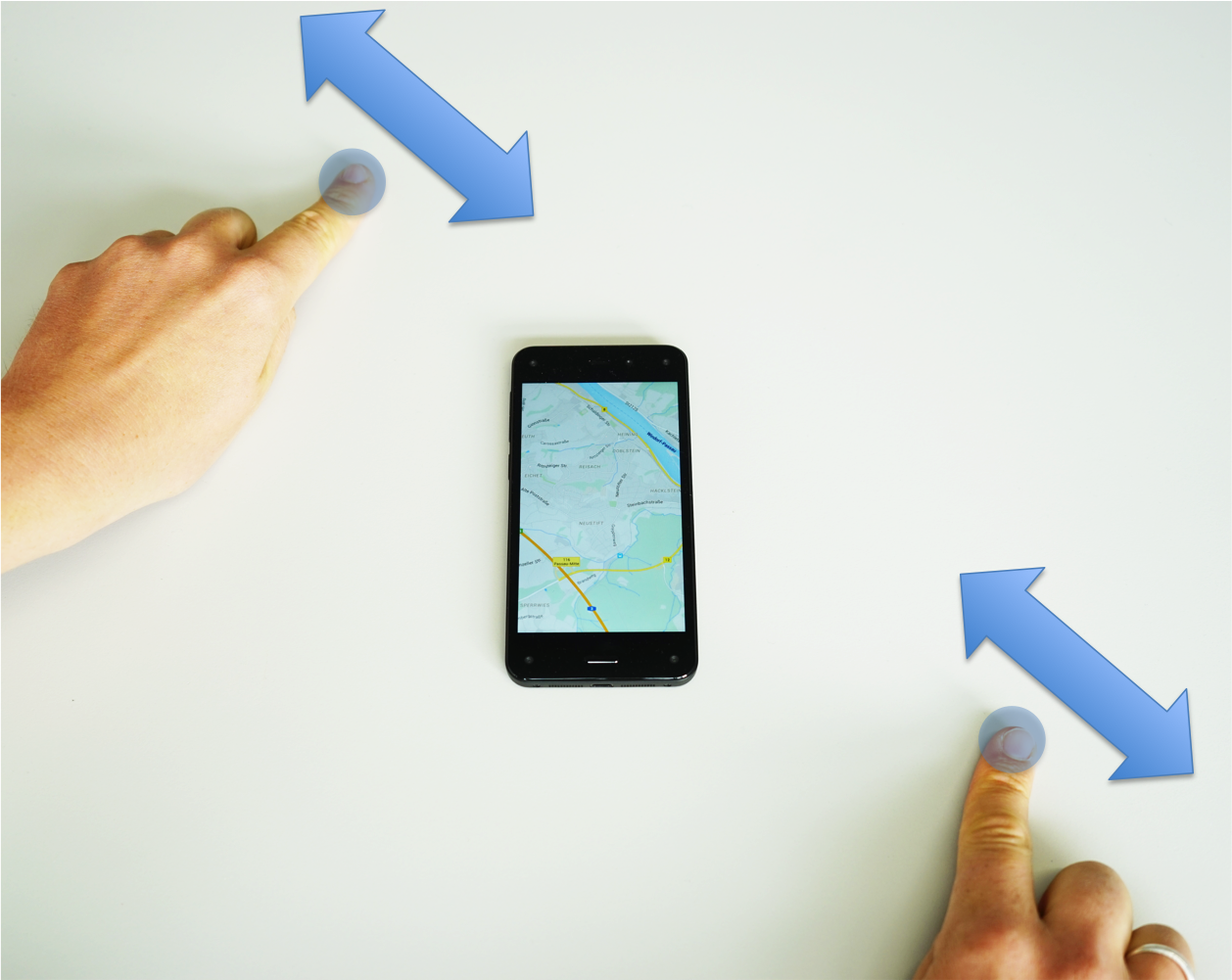}\\
    \includegraphics[width=0.9\marginparwidth]{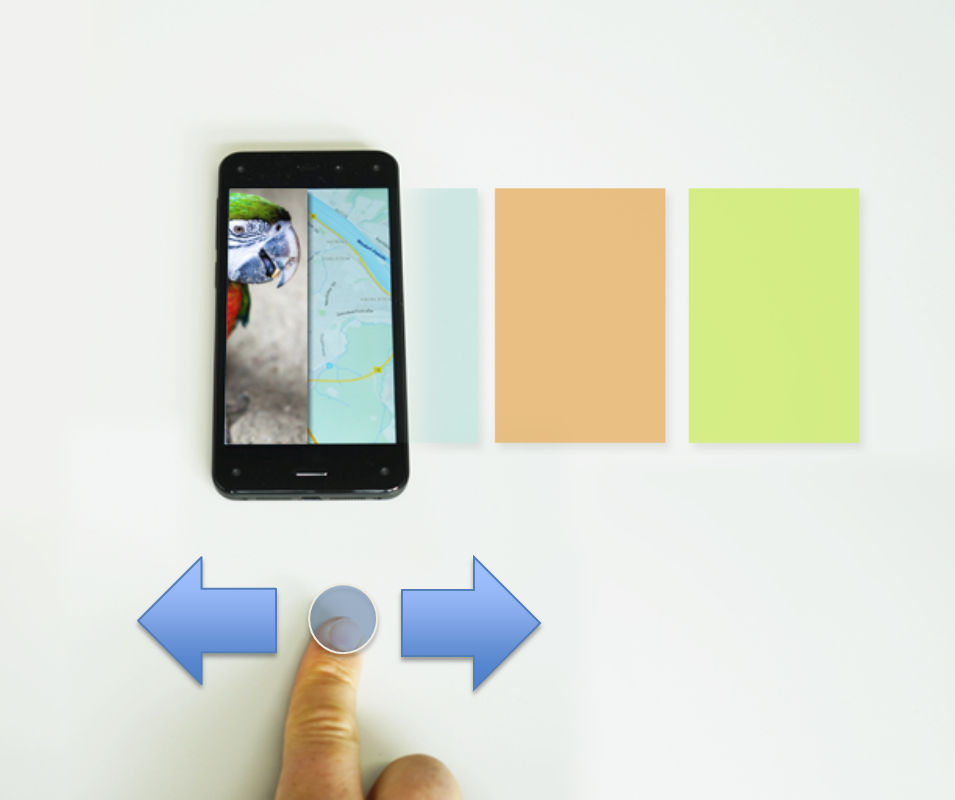}\\
    \caption{Top: Panning and zooming a map requires fewer repetitive
gestures, compared to touch-screen only interaction and avoids screen
occlusions. Bottom: A virtual ribbon allows browsing open applications, by panning outside the screen.}~\label{fig:apps}
  \end{minipage}
\end{marginfigure}

\subsection{Map Navigation}
Touch-based map navigation on mobiles is limited by the small screen space: Panning to distant objects 
or zooming large distances typically involves repeated drag and pinch gestures. Wearing gloves, for example while skying or biking, may prevent using touch to interact with the phone. 
Using GlassHands, one can pan and zoom utilizing the surface, or the air around the device (see Figure~\ref{fig:apps}) or simultaneously pan and trace over an item of interest (see Figure~\ref{fig:teaser}, right). 

Moreover, navigating a map by outside device gestures avoids occlusion by the fingers, which is an advantage for small displays. Touching solely inside or outside the display causes the map to be paned and zoomed normally.

\subsection{Task Switching}
Switching tasks on mobiles involves typically at least two touches and an additional pan action, searching for the requested application among a list of prior used applications. 
Furthermore, if task switching includes cutting and pasting items via a pop-up menu, at least six consecutive touch actions are needed. 

We allow browsing application using a linear ribbon metaphor (see Figure \ref{fig:apps}) and enable fast cutting and pasting: The user selects an item to be moved, and holds it. Next, the user switches to the target application, either directly through tapping on locations as mentioned above or by panning the ribbon with a finger drag outside the phone. Upon reaching the target application, the user releases the item to paste it into the target. Prior to releasing, the user can place the item at the desired location within the target application by dragging it on the phone's display. This operation can also be done in mid-air, where the holding hand thumb is used for item selection.

\subsection{Mid-Air Music Player}



Working with touch screens of mobile devices while wearing gloves, such as while skiing, viewing a map on a motorbike (at a traffic stop) or answering a phone while using work gloves, can be cumbersome. Users have to take off their gloves to operate common capacitive touch screens, e.g., when browsing through a music collection or when unlocking the screen. While touch screen capable gloves exist, they are definitely rarer than ski goggles. Furthermore, touching mobile screens with gloves can lead to an amplified fat finger problem. We have implemented a music player application that allows browsing music collections using mid-air gestures. The user can initiate a gesture by holding the hand next to the phone. Then, hand movements to either side of the phone are mapped to a scrolling list, as can be seen in Figure~\ref{fig:midAir}. The large available interaction space allows fine accuracy of selection. The same application can also be used with hand gestures on surfaces. 

\section{Discussion and Conclusion}

We demonstrated that interaction at the periphery of a mobile device is feasible given a front-facing RGB camera and a user wearing sunglasses, or any other reflector. In contrast to previous work \cite{song2014air}, we significantly broaden the interaction space around an unmodified mobile device. In contrast to Surround-See~\cite{yang2013surround}, we enable large interaction space, without the need for special hardware, just everyday common apparel, and with keeping the phone thin and small. Removing the need for special hardware enable us to easily deploy this solution, such as through a store application.

The proposed system has several limitations, some which may be addressed in future work:

An inherent limitation of our approach is that users have to wear reflective glasses, which have to be visible in the field of view of the front camera (typically 70-80 degrees field of view). 
Dark glasses allows the camera to view a clear reflection without the view of the user eyes behind them. If the user face is relatively dark, such as in the case of a desk lamp illuminating the table, while the user's head stays in the dark, then regular glasses can be used just as well.

In general, the usage of dark glasses in low-light environments, like indoor offices, might not be appropriate, both due to perceptual and social reasons. However, there are many situations, both indoors and outdoors, where wearing reflective eye-wear is common: workers wearing safety goggles, skiers, divers, motorcyclists and so on.

On-surface interactions as described above enable comfortable work, where the users hands are supported by the surface, and the tap of a finger on the surface can be used to detect touch. However, the technology is not limited to surface-based interaction. The same 2D interaction may be used in mid-air around the phone with ample gestures. Such in-air interaction with Glasshands could help skiers operate their mobiles without the need to take off their gloves. 

\begin{marginfigure}[-15pc]
  \begin{minipage}{\marginparwidth}
    \centering
    \includegraphics[width=0.9\marginparwidth]{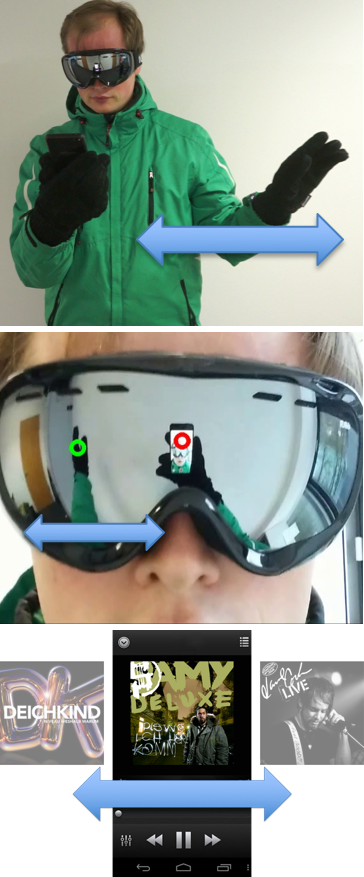}
    \caption{Top: Mid-air sliding gesture. GlassHands support large hand gestures, that enables continues scrolling or selection from a large list of options. Middle: Close-up with detected phone (red circle) and glove (green circle). Bottom: The album selection of a music player is operated with left and right sliding gestures.}
\label{fig:midAir}
  \end{minipage}
\end{marginfigure}

Also, GlassHands could be used by bikers, who have mounted their phone on the handle bar.  Sliding the hands along the handle bar, relatively far from the phone, could be utilized to steer on-screen applications (see Figure~\ref{fig:midAir}, right). A possible application may sense the direction of a pointed hand, reflected in the helmet visor, and use the phone GPS and orientation to announce the street number of the house the biker is pointing at.
Divers could use GlassHands to interact with their phone, stored inside a watertight casing. 
In a similar fashion, workers who wear protective glasses and gloves may interact using large gestures around the phones.

Measuring the hand positioning accuracy shows that a minimal distance of at least 5 cm is needed to reliably separate two individual touch down events on the surface, when a single frame is used for measurement. It is possible to increase the measurement accuracy, for example, using temporal filtering.

%

Furthermore, we deliberately employed simple and efficient computer vision techniques throughout our pipeline to demonstrate the feasibility of the GlassHands concept. These simple algorithms can not cope well with complex environments, typically found in real-world situations. For practical implementations in commercial applications, more robust algorithms should be used (see also below). 

At the camera resolution we used (0.9  megapixel), the image of each lens is about 130 by 170 pixels, which limits the maximum accuracy of the hands location to about 0.5 cm. Better front cameras are already available and should improve the quality of detection and location estimation. 

\begin{marginfigure}[5pc]
  \begin{minipage}{\marginparwidth}
    \centering
    \includegraphics[width=0.9\marginparwidth]{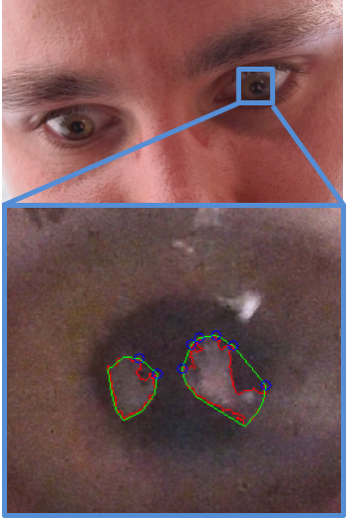}
    \caption{Top: 6 megapixel image of a users face area. Bottom: Hands are detected in the cropped corneal area (0.16 megapixel).}
\label{fig:eye}
  \end{minipage}
\end{marginfigure}

In the future, we want to support a wider variety of glasses models with different reflection and curvature properties.
Moreover, the use of reflections directly of the user's eye using corneal imaging~\cite{nitschke2013corneal} could overcome the need for eye-wear. Figure~\ref{fig:eye} shows the reflection of the user eye, as captured by a 6 megapixel DSLR 
placed on a table at the same distance from the user as the phone. Hands are detected using a standard hand tracking approach in the cornea of the user. Currently available mobile phones do not have such high resolution front cameras, but future models may support them, making the presented technique fit more usage scenarios and remove the need of wearing reflective glasses.

Furthermore, by using both lenses of the glasses, one could model two catadioptric camera systems and apply stereo techniques to recover depth values. We plan to investigate how to estimate such camera models on the fly. Depending on the quality of the stereo reconstruction, usage may range from determining the height of the user's hands above the surface to possible replacement of body-worn depth sensors.

GlassHands allows to sense an ample area in a phone's vicinity. There are options to recognize objects in space and react accordingly. The interaction may involve everyday objects, such as toys on a table, a board-game, or ingredients on a kitchen counter. As a next steps, our approach could be extended to ad-hoc spatial registration of multi-display environments without the need for external cameras \cite{radle2014huddlelamp}.

We believe that new sensing capabilities for phones will help spreading spatially aware applications. In many cases, the development of such applications is hampered by the limited distribution of required hardware. Hardware manufacturers, on the other hand, may hesitate to include new technology without proven value for applications~\cite{buxton2008long}. We hope that an approach such as GlassHands may break this circle by enabling applications aimed at a specific scenario, such as Skiing, to be commercially successful using existing hardware. Ultimately, this may encourage the development of new dedicated sensors integrated in consumer phones.




\balance{} 

\bibliographystyle{SIGCHI-Reference-Format}
\bibliography{gh}

\end{document}